\documentclass[%
reprint,
superscriptaddress,
 amsmath,amssymb,
 aps,
 prl,
floatfix,
]{revtex4-1}

\usepackage{commath}
\usepackage{esdiff} 
\usepackage{nicefrac}
\usepackage{graphicx}
\usepackage[caption=false]{subfig}
\usepackage{dcolumn}
\usepackage{bm}
\usepackage[colorlinks]{hyperref}
\usepackage{xcolor}
\usepackage{siunitx} 
\usepackage[utf8]{inputenc}
\usepackage[capitalise,nameinlink]{cleveref} 
\Crefname{equation}{Eq.}{Eqs.}
\Crefname{figure}{Fig.}{Figs.}
\Crefname{tabular}{Tab.}{Tabs.}

\DeclareSIUnit\parsec{pc}
\DeclareSIUnit\efolds{e\text{-}folds}
\DeclareSIUnit\planckmass{m_\mathrm{p}}
\DeclareSIUnit\clight{\mathit{c}}

\begin{document}


\title{A case for kinetically dominated initial conditions for inflation}

\author{L.~T.~Hergt}
 \email{lh561@mrao.cam.ac.uk}
\author{W.~J.~Handley}%
 \email{wh260@mrao.cam.ac.uk}
\affiliation{%
 Astrophysics Group, Cavendish Laboratory, J.~J.~Thomson Avenue, Cambridge, CB3~0HE, UK
}%
\affiliation{%
 Kavli Institute for Cosmology, Madingley Road, Cambridge, CB3~0HA, UK
}%
\author{M.~P.~Hobson}%
 \email{mph@mrao.cam.ac.uk}
\affiliation{%
 Astrophysics Group, Cavendish Laboratory, J.~J.~Thomson Avenue, Cambridge, CB3~0HE, UK
}%
\author{A.~N.~Lasenby}%
 \email{a.n.lasenby@mrao.cam.ac.uk}
\affiliation{%
 Astrophysics Group, Cavendish Laboratory, J.~J.~Thomson Avenue, Cambridge, CB3~0HE, UK
}%
\affiliation{%
 Kavli Institute for Cosmology, Madingley Road, Cambridge, CB3~0HA, UK
}%

\date{\today}

\begin{abstract}
    We make a case for setting initial conditions for inflation at the Planck epoch in the kinetically dominated regime. For inflationary potentials with a plateau or a hill, i.e.\ potentials that are bounded from above within a certain region of interest, we cannot claim complete ignorance of the energy distribution between kinetic and potential energy, and equipartition of energy at the Planck epoch becomes questionable. We analyse different classes of potentials in phase-space and quantify the fraction of the Planck surface that is kinetically dominated. For the small amplitudes of the potentials as suggested by current data, the Planck surface lies in the region of kinetic dominance for almost all values of interest of the inflaton field.
\end{abstract}

\maketitle

\section{Introduction}

{\NoHyper\citeauthor{Handley2014}\endNoHyper}~\citep{Handley2014,KineticNote} show under broad assumptions that classical inflationary universes generically emerge in a regime where the kinetic energy of the inflaton dominates over the potential. In contrast, the traditional procedure for setting initial conditions for inflation defines them at the Planck epoch with a total energy density of the order of $\rho\sim\si{\planckmass\tothe4}$~\cite{Belinsky1985,Linde1985,Belinsky1988,Linde2007} and, lacking any further prior constraints, partitions inflaton energy equally between kinetic and potential energy~\cite{Linde1985,Linde2007,Belinsky1988,Boyanovsky2006a,Destri2008}.

In this letter we show how the choice of certain potentials gives additional prior constraints. We make a case for why initial conditions at the Planck time should be set using kinetic dominance as opposed to assuming that the potential holds half the energy.

\section{The Planck surface in phase space}

Energies beyond the Planck scale require a quantum theory of gravity. Only when $\rho\lesssim\si{\planckmass\tothe4}$ may we set initial conditions for any \emph{classical} evolution~\cite{Linde1985,Belinsky1985,Belinsky1988,Linde2007}. For the inflaton field~$\phi$ in a spatially-flat universe this means starting from the Planck circle where, neglecting spatial inhomogeneities:
\begin{alignat}{3}
	\label{eq:plancksurface}
	&\frac{1}{2} \dot\phi^2 &&+ V(\phi) &&= \si{\planckmass\tothe4} , \\
    \label{eq:planckcircle}
    &y^2 &&+ x^2 &&= r^2 ,
\end{alignat}
with $x \equiv \mathrm{sgn}(\phi) \sqrt{V(\phi)}$ and $y \equiv \dot\phi / \sqrt{2}$.
In the $(x, y)$~parametrisation this is a circle with radius $r=\si{\planckmass\squared}$. However, in the $(\phi, \dot\phi)$~phase-space the Planck surface is not in general circular. Depending on the potential~$V(\phi)$, the shape of the surface in the $(\phi,\dot\phi)$~phase-space changes, leading to features such as local maxima or asymptotes in~$\dot\phi$. In the $(x, y)$~parametrisation this manifests as part of the circle becoming effectively excluded.

\begin{figure}[b]
		\centering
		\includegraphics[width=0.95\columnwidth]{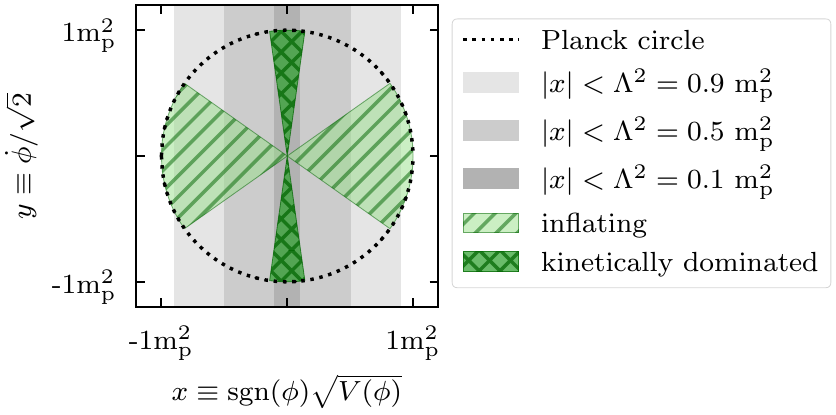}
        \caption{The Planck circle in the $(x, y)$~parametrisation is depicted by the black dotted line. The stripy and checked hatched regions correspond to the inflating region and the kinetically dominated region respectively (see \cref{eq:inflating,eq:kd}). The grey shading highlights the effective $x$-range relevant for three given values of~$\Lambda^2$ for potentials such as plateau or hilltop potentials (e.g.\ the Starobinsky and Landau-Ginzburg potentials defined in \cref{eq:starobinsky,eq:hilltop} respectively).}
    \label{fig:planckcircle}
\end{figure}

We define the inflating region via $\ddot{a}>0$, or equivalently:
\begin{equation}
	\label{eq:inflating}
	\dot\phi^2<V(\phi) .
\end{equation}
Similarly, we conservatively define the region of kinetic dominance (KD) by the condition that kinetic energy dominates the potential by two orders of magnitude:
\begin{equation}
	\label{eq:kd}
	\dot\phi^2 > 100~V(\phi) .
\end{equation}
\Cref{fig:planckcircle} shows the Planck circle in the $(x, y)$~parametrisation as a black dotted line. The inflating and KD regions are sectors centred along the $x$- and $y$-axis respectively. This picture holds irrespective of the choice of potential~$V(\phi)$. However, the choice of potential can effectively exclude parts of this circle: e.g.\ for the plateau and hilltop potentials defined in \cref{eq:starobinsky,eq:hilltop} below, this is illustrated by the grey shaded regions covering smaller values of~$x$ as the amplitude parameter~$\Lambda$ decreases. Note how for very small~$\Lambda^2$ the relevant part of the circle is reduced to within the KD region~\cite{Mishra2018}.
In \cref{fig:phasespace}, discussed more fully below, we see that the picture changes significantly upon changing to $(\phi, \dot\phi)$~phase-space and for alternative potentials. 
In the following section we review our three representative potential classes: quadratic, plateau and hilltop.

\section{Inflaton potentials}

\subsection{Quadratic potential}

The simplest potential typically used for single field inflation models is given by:
\begin{equation}
	\label{eq:m2phi2}
	V(\phi) = m^2 \phi^2 ,
\end{equation}
where~$m$ is the mass of the inflaton field. This quadratic potential is often defined with a multiplicative factor~$\frac{1}{2}$, which is irrelevant for this analysis and omitted here for reasons of compatibility with other power law potentials. In general, $m\sim$~\SI{e-6}{\planckmass} in order to produce an appropriate primordial power spectrum amplitude~\cite{Planck2015parameters}. Though quadratic potentials are disfavoured by current data because of their excess production of gravitational waves~\cite{Planck2015inflation}, we include them for comparison with other models and traditional theoretical discussion.

The quadratic potential plays a special role in the phase-space representation because $(x, y)$ is equivalent to $(\phi, \dot\phi)$~phase-space up to a factor of~$m$. Because of this close relationship for quadratic potentials, the distinction between the two spaces is often overlooked. In the following sections we discuss the effects of considering alternative potentials.

\subsection{Plateau potentials}

Some high-energy models give rise to plateau-like potentials~\cite{Starobinsky1980,Dvali1998,Cicoli2009}. A popular example is the Starobinsky potential~\cite{Starobinsky1980}:
\begin{equation}
	\label{eq:starobinsky}
	V(\phi) = \Lambda^4 \left[ 1 - e^{-\sqrt{2/3} \phi / \si{\planckmass}} \right]^2 , 
\end{equation}
where the scale $\Lambda^2\sim$ \SI{e-6}{\planckmass\squared} is comparable to the inflaton mass~$m$ of the quadratic potential. Unlike the quadratic potential, due to concavity the plateau potential produces very few gravitational waves making it preferred by Planck data~\cite{Planck2015inflation}. 

For small amplitudes, $\Lambda<\si{\planckmass}$, the asymptote in the potential for $\phi\rightarrow\infty$ translates through to $(\phi, \dot\phi)$~phase-space, where~$\dot\phi$ converges to $\sqrt{2(\si{\smash{\planckmass\tothe4}}-\Lambda^4)}$ from its global maximum at \smash{$\SI[parse-numbers=false]{\sqrt{2}}{\planckmass\squared}$}. In $(x, y)$~space this asymptote manifests as a cut from the Planck circle as an unattainable region. No value of $\phi>0$ can reach the region \smash{$x=\sqrt{V}\ge\Lambda^2$}. Note that on the exponentially growing side $\phi<0$, the potential and thus~$x$ increase beyond the Planck scale irrespective of the amplitude~$\Lambda$. However, for cosmic inflation only the plateau side of the potential is of interest to us. 

\subsection{Hilltop potentials}

The double-well (Landau-Ginzburg) potential is given by:
\begin{equation}
	\label{eq:hilltop}
	V(\phi) = \Lambda^4 \left[ 1 - \left( \frac{\phi}{\mu} \right)^2 \right]^2 .
\end{equation}
This potential has been frequently studied in the context of cosmic inflation and spontaneous symmetry breaking~\cite{Zee1979,Accetta1985,Lucchin1986,Kaiser1995,Bezrukov2007,Cerioni2009,Burns2016,Bostan2018}. It is a particular realisation of the family of quadratic hilltop potentials~\cite{Boubekeur2005a} that ensures positivity (required here for the calculation of the Planck surface). As for plateau potentials, the concavity in the hilltop region ensures low production of primordial gravitational waves~\cite{Planck2015inflation}.

The inflaton rolls away from a local maximum at $\phi=0$ to a minimum at~$\phi=\pm\mu$. We can identify the region of interest $-\mu<\phi<\mu$ where the potential is bounded from above ($|x|=\sqrt{V}\le\Lambda^2$), which effectively cuts the Planck circle defined by \cref{eq:planckcircle} and shown in \cref{fig:planckcircle}.

\section{Inflaton Phase-Space}

\begin{figure*}[p]
	\centering
	\includegraphics[width=\textwidth]{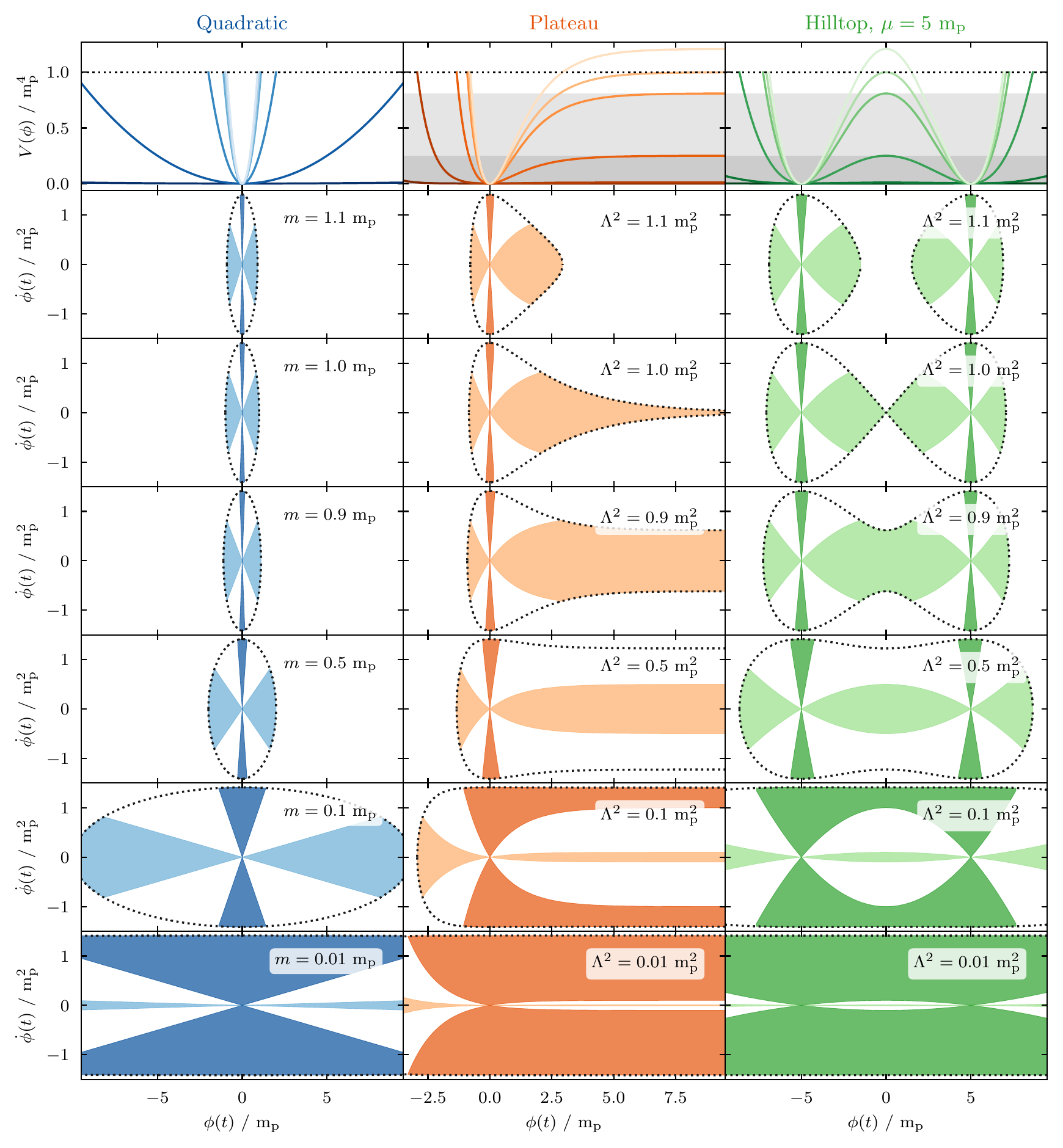}
    \caption{\label{fig:phasespace}The top row shows $V(\phi)$ for the quadratic potential from \cref{eq:m2phi2} in blue (left), for the plateau potential from \cref{eq:starobinsky} in orange (middle), and for the hilltop potential from \cref{eq:hilltop} in green (right). The lines in each potential plot grow darker with decreasing~$m$ or~$\Lambda^2$. The grey shaded regions correspond to those in \cref{fig:planckcircle} delimiting the maximum value of the plateau or hill region for selected values of~$\Lambda^2$. The other panels depict $(\phi, \dot\phi)$~phase-space diagrams for the corresponding potentials with values for~$m$ and~$\Lambda^2$ decreasing from the second row downwards.  The enveloping dotted line denotes the Planck surface. Light shading corresponds to the inflating regions where $\dot\phi^2<V(\phi)$. Dark shading corresponds to the kinetically dominated regions where $\dot\phi^2>100~V(\phi)$. Note that realistic values for~$m$ and~$\Lambda^2$, i.e.\ conforming with data for the primordial amplitude~$A_\mathrm{s}$ from e.g.\ the Planck satellite~\cite{Planck2015parameters}, are orders of magnitude smaller (i.e.\ continuing the rows further down) than the values picked here for reasons of visualisation.}
\end{figure*}

In \cref{fig:phasespace} we plot the quadratic, plateau, and hilltop models side by side. The top row shows the potentials from \cref{eq:m2phi2,eq:starobinsky,eq:hilltop}. The other panels are $(\phi, \dot\phi)$~phase-space plots for decreasing values of the inflaton mass~$m$, or amplitude~$\Lambda$ from top to bottom. In analogy with \cref{fig:planckcircle}, the Planck surface is given as a black dotted line and the shaded regions correspond to the inflating and kinetically dominated regions (light and dark shading respectively).

For the quadratic potential, the shape of the Planck surface is ellipsoidal and thus closed. For the plateau potential this ellipse is drawn out to $\phi\rightarrow\infty$ because of the asymptote in the potential $V(\phi\rightarrow\infty)=\Lambda^4$. Only for $\Lambda>$ \si{\planckmass} does the Planck surface become closed. The double-well potential consists of two such deformed ellipsoids which are connected for $\Lambda\le$ \si{\planckmass}.

Of particular interest is the kinetically dominated (KD) regime that only covers a tiny part at the poles of the Planck surface when~$m$ or~$\Lambda$ are large. When decreasing~$m$ in the quadratic model, the KD regime expands in proportion to the Planck surface. For the plateau model, the KD region can continue to stretch while the Planck surface already extends to infinity. Analogously, in the hilltop case the Planck surface already spans from~$-\mu$ to~$\mu$ and KD eventually spreads to cover the whole range for very small~$\Lambda$.

\section{Priors on initial conditions}

\begin{figure*}[p]
	\centering
    \includegraphics[width=\textwidth]{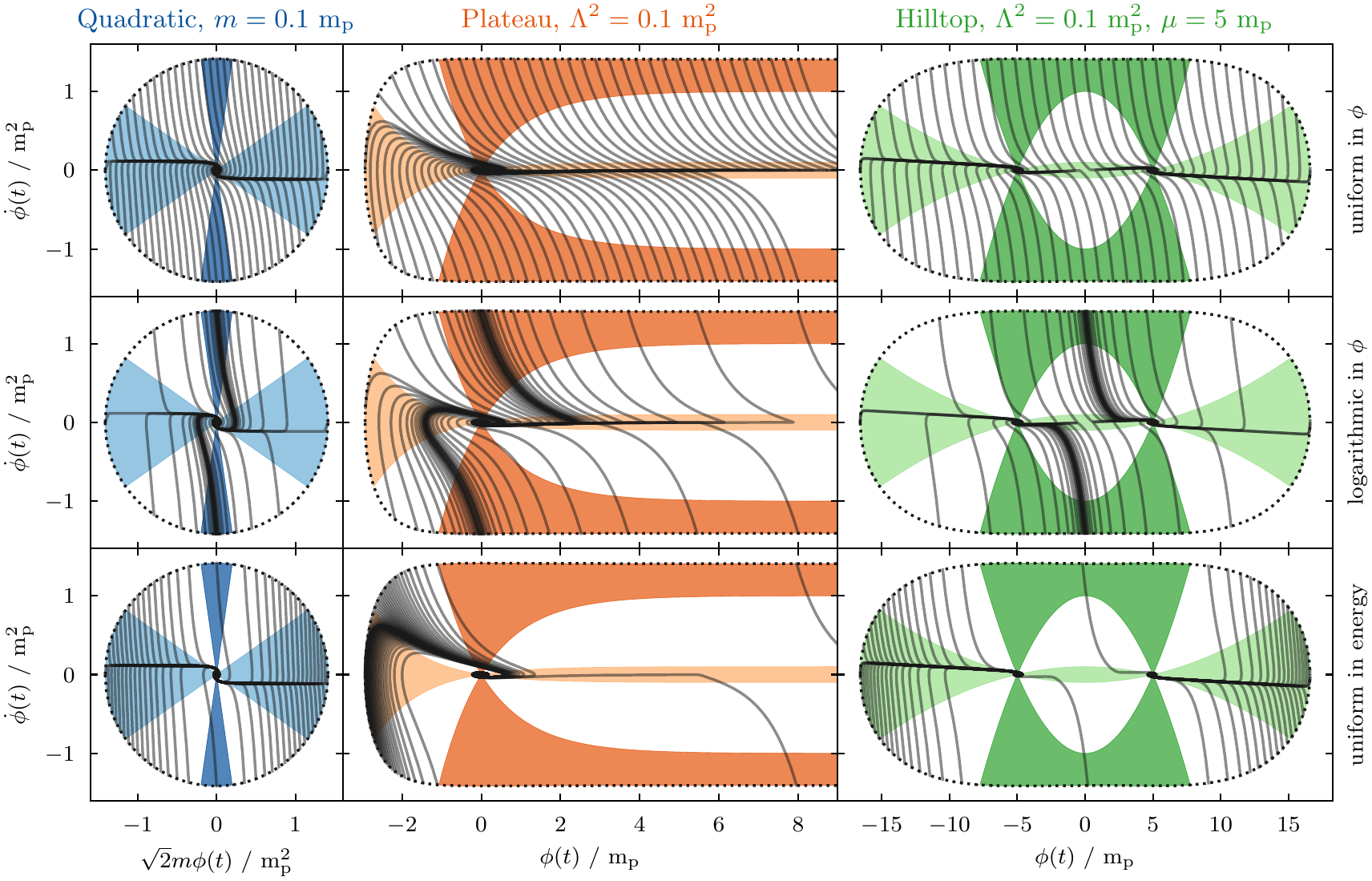}
    \caption{\label{fig:priors}$(\phi, \dot\phi)$~phase-space diagrams with inflaton trajectories for the quadratic potential in blue (left), the plateau potential in orange (middle), and the hilltop potential in green (right). Light shading correspond to the inflating regions where $\dot\phi^2<V(\phi)$. Dark shading corresponds to the kinetically dominated regions where $\dot\phi^2>100~V(\phi)$. The enveloping dotted line denotes the Planck surface. In the top row initial conditions were set with a uniform prior on~$\phi$, with a logarithmic prior on~$\phi$ in the middle row, and with a uniform prior on the potential energy in the range $[0,\si{\planckmass\tothe4}]$ in the bottom row.}
    \vskip 0.8\baselineskip
    \includegraphics[width=\textwidth]{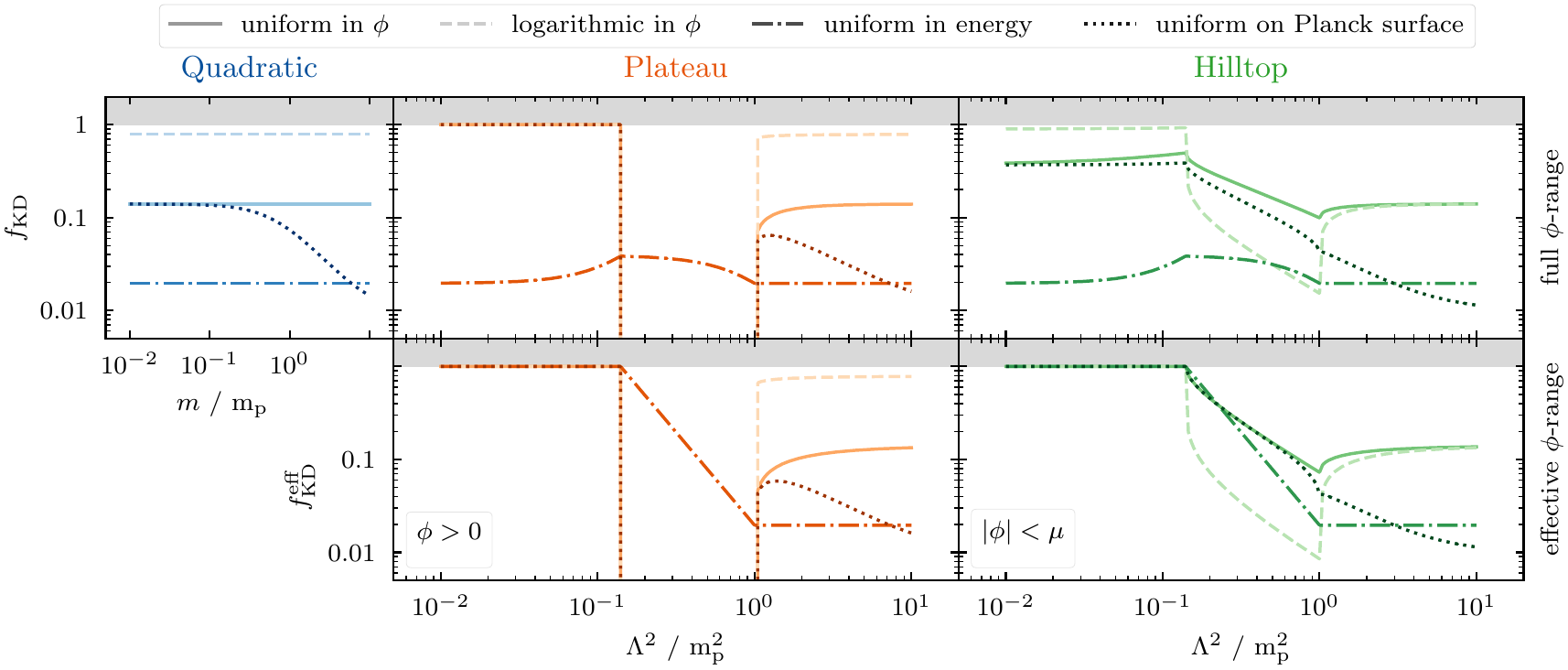}
    \caption{\label{fig:kdfractions}Fraction~$f_\mathrm{KD}$ of the kinetically dominated region to the full Planck region in $(\phi, \dot\phi)$~phase-space corresponding to the prior choices from \cref{fig:priors}. The second row shows how these fractions change when we only consider the $\phi$-range effectively necessary for the type of inflation we are interested in, i.e.\ $\phi>0$ for the plateau potential and $|\phi|<\mu$ in the hilltop case.}
\end{figure*}

When setting initial conditions at the Planck scale, the traditional approach is to assume equipartition between kinetic and potential energy~\cite{Linde1985,Linde2007,Belinsky1988,Boyanovsky2006a,Destri2008}. It then is concluded that an initial value for the inflaton~$\phi_0$ may take any value respecting the condition that $V(\phi_0)\lesssim\si{\planckmass\tothe4}$. With only this constraint, for very small inflaton masses $m\sim\SI{e-5}{\planckmass}$ this means that $\phi_0$ is allowed to take a huge range of values spanning many orders of magnitude. Faced with complete ignorance regarding the scale of~$\phi_0$ it is arguably more reasonable to choose a logarithmic prior for the initial value of the inflaton field.

In \cref{fig:priors} the trajectories of the inflaton field in $(\phi,\dot\phi)$~phase-space are plotted using different priors for the generation of initial conditions on the Planck surface. \Cref{fig:kdfractions} shows the fraction of trajectories starting in the kinetically dominated (KD) region for each prior respectively. We compare uniform and logarithmic priors in~$\phi$, a uniform prior on the energy distribution between potential and kinetic energy, and a prior uniform on arc-length on the Planck surface. In general we can see that equipartition of energy pushes the trajectories outside the KD window whereas a scale-invariant logarithmic prior has the opposite effect. 

For the quadratic potential, prior choice plays a significant role. The KD fractions for priors in~$\phi$ or energy are independent of the inflaton mass~$m$. Only for the uniform prior on arc-length do we find a decrease of the KD fraction at large mass, when the Planck surface spans a very small $\phi$-range.
For energy equipartition we get a KD fraction of only $f_\mathrm{KD}\approx\SI{2}{\percent}$ and for a uniform prior on~$\phi$ about \SI{14}{\percent}. A logarithmic prior on~$\phi$ raises the fraction to about \SI{79}{\percent}, where we start at $\phi_\mathrm{start}=\SI{e-4}{\planckmass\squared}/m$, which is roughly the scale from which a sufficient amount of \si{\efolds} of inflation is produced for realistic values of~$m$.

Unlike the quadratic potential, the KD fraction for the plateau and hilltop potentials does depend on~$\Lambda^2$. This is related to the changing shape of the Planck surface illustrated in \cref{fig:phasespace} above. For $\Lambda>\si{\planckmass}$ the Planck surface is bounded and the KD fractions display a similar behaviour to the quadratic model. The fraction drops towards smaller values of~$\Lambda$, as the Planck surface stretches across a greater $\phi$-range, whereas the KD regime does not. Crossing $\Lambda=\si{\planckmass}$ reverses this behaviour in the hilltop case. As the Planck surface spans across $\phi=0$, while the KD regime continues to spread out, the KD fraction starts to rise. Finally for $\Lambda^2<\si{\planckmass\squared}/\sqrt{51}$, the KD fraction levels off. For the Starobinsky model, where the Planck surface stretches to infinity, the fraction stays at zero at intermediate amplitudes, and jumps to unity once the KD regime also spreads to infinity.

For both the plateau and hilltop potentials we note how a uniform prior on the energy distribution causes only a small number of trajectories to start in the regions of interest: $\phi>0$ and $-\mu<\phi<\mu$ respectively. Once we restrict ourselves to the $\phi$-ranges of importance, all trajectories start in the KD region irrespective of the prior choice. This change from the full $\phi$-range to a restricted $\phi$-range is illustrated by going from the first to the second row in \cref{fig:kdfractions}. Thus, for potentials with an upper bound in the region of interest (i.e.\ most concave models) initial conditions for inflation should naturally be set in the KD regime.

\section{Conclusion}

We have shown how phase-space considerations can be significantly more complicated in comparison with the simplistic Planck circle picture when considering models other than quadratic inflation. This is particularly relevant for potentials with features such as plateaus or hills, i.e.\ concave and bounded from above in the region of interest for slow-roll inflation. With a bound on potential energy we cannot use energy equipartition as a prior at the Planck epoch. Instead, we claim that initial conditions for inflation at the Planck epoch should be set in the kinetically dominated regime.

\begin{acknowledgments}
LTH would like to thank the Isaac Newton Trust and the STFC for their support. WJH was supported by a Gonville~\& Caius Research Fellowship.
\end{acknowledgments}

\bibliography{KineticDominanceBib,CollaborationBib,footnote}

\end{document}